\documentclass[conference]{IEEEtran}
\IEEEoverridecommandlockouts
\usepackage{cite}
\usepackage{amsmath,amssymb,amsfonts}
\usepackage{algorithmic}
\usepackage{graphicx}
\usepackage{textcomp}
\def\BibTeX{{\rm B\kern-.05em{\sc i\kern-.025em b}\kern-.08em
		T\kern-.1667em\lower.7ex\hbox{E}\kern-.125emX}}

\usepackage{times}
\usepackage{epsfig}

\usepackage{afterpage}
\usepackage{booktabs}
\usepackage{multirow}
\usepackage{siunitx}
\usepackage{subcaption}

\usepackage[pagebackref=false,breaklinks=true,letterpaper=true,colorlinks,bookmarks=false]{hyperref}

\usepackage{xspace}
\makeatletter
\DeclareRobustCommand\onedot{\futurelet\@let@token\@onedot}
\def\@onedot{\ifx\@let@token.\else.\null\fi\xspace}

\def\eg{\emph{e.g}\onedot} 

\def\ie{\emph{i.e}\onedot}

\def\etal{\emph{et al}\onedot}
\makeatother

\newcommand{\stufig}[5]                                       % images with specified placement
{
	\begin{figure}[#5]
		\begin{center}
			\includegraphics[#1]{#2}
			\vspace{.2cm}
			\caption{#3}
			\label{#4}
		\end{center}
		\vspace{-\baselineskip}
	\end{figure}
}

\newcommand{\stufigstar}[5]                                   % full-width images with specified placement
{
	\begin{figure*}[#5]
		\begin{center}
			\includegraphics[#1]{#2}
			\caption{#3}
			\label{#4}
		\end{center}
		\vspace{-1.5\baselineskip}
	\end{figure*}
}

\newenvironment{stusubfig}[1]
{
	\begin{figure}[#1]
		\begin{center}
		}
		{
		\end{center}
	\end{figure}
}

\newenvironment{stusubfig*}[1]
{
	\begin{figure*}[#1]
		\begin{center}
		}
		{
		\end{center}
	\end{figure*}
}

\DeclareMathOperator*{\argmin}{argmin}

\newcommand\colvec[2]{\left[ \begin{array}{c} #1 \\ #2 \end{array} \right]}

\newcolumntype{H}{>{\setbox0=\hbox\bgroup}c<{\egroup}@{}} % Column type that hides the content

\makeatletter
\def\maketag@@@#1{\hbox{\m@th\normalfont\normalsize#1}}
\makeatother

\usepackage{xcolor}

\begin{document}
	
	\title{R$^3$SGM: Real-time Raster-Respecting Semi-Global Matching for Power-Constrained Systems}

	\author{
		\begin{tabular}{c@{\hskip 0.85cm}c@{\hskip 0.85cm}c@{\hskip 0.85cm}c@{\hskip 0.85cm}c}
			Oscar Rahnama & Tommaso Cavallari$^\star$ & Stuart Golodetz$^\star$ & Simon Walker & Philip H.\ S.\ Torr
		\end{tabular}%
		\thanks{$^\star$ TC and SG assert joint second authorship.}
		\thanks{{\texttt{\{oscar.rahnama,tommaso.cavallari,stuart\}@five.ai}}}
		\thanks{OR is with both the University of Oxford and FiveAI Ltd. TC and SG were with the University of Oxford, and are now with FiveAI Ltd. SW is with FiveAI Ltd. PT is with the University of Oxford.}
	}

	\maketitle
	
%%%%%%%%% ABSTRACT
\begin{abstract}
    \noindent Stereo depth estimation is used for many computer vision
applications. Though many popular methods strive solely for depth quality, for
real-time mobile applications (e.g.\ prosthetic glasses or micro-UAVs), speed
and power efficiency are equally, if not more, important. Many real-world
systems rely on Semi-Global Matching (SGM) to achieve a good accuracy vs.\ speed
balance, but power efficiency is hard to achieve with conventional hardware,
making the use of embedded devices such as FPGAs attractive for low-power
applications. However, the full SGM algorithm is ill-suited to deployment on
FPGAs, and so most FPGA variants of it are partial, at the expense of accuracy.
In a non-FPGA context, the accuracy of SGM has been improved by More Global
Matching (MGM), which also helps tackle the streaking artifacts that afflict
SGM. In this paper, we propose a novel, resource-efficient method that is
inspired by MGM's techniques for improving depth quality, but which can be
implemented to run in real time on a low-power FPGA. Through evaluation on
multiple datasets (KITTI and Middlebury), we show that in comparison to other
real-time capable stereo approaches, we can achieve a state-of-the-art balance
between accuracy, power efficiency and speed, making our approach highly
desirable for use in real-time systems with limited power.
	
\end{abstract}

%\vspace{-\baselineskip}

%-------------------------------------------------------------------------
\section{Introduction}

\noindent Numerous computer vision applications, including 3D voxel scene
reconstruction \cite{prisacariu2017, golodetz2018arxiv}, object recognition
\cite{lai2013}, 6D camera relocalisation \cite{shotton2013, cavallari2017}, and
autonomous navigation \cite{oleynikova2015reactive, hicks_depth-based_2013},
either rely on, or can benefit from, the availability of depth to capture 3D
scene structure. Active approaches for acquiring depth,
based on structured light \cite{zhang2012kinect} or LiDAR, produce high-quality
results. However, the former performs poorly outdoors, where sunlight washes out
the infrared patterns it uses, whereas the latter is generally expensive and
power-hungry, whilst simultaneously only producing sparse depth. Significant
attention has thus been devoted to passive methods of obtaining dense depth from
either monocular or stereo images.
Although recent approaches based on deep learning \cite{liu2016,godard2017} have
made progress in the area, monocular approaches, which only require a single
camera, struggle in determining scale \cite{tateno2017}. Stereo approaches, as a
result, are often preferred when multiple cameras can be used, with binocular
stereo methods (which achieve a compromise between quality and cost) proving
particularly popular.

Many binocular stereo methods estimate disparity by finding correspondences between the two images.
They typically involve four phases~\cite{scharstein2002taxonomy}: (a) matching cost computation, (b) cost
aggregation, (c) disparity optimisation, and (d) disparity refinement. At a high
level, such methods can be classified into two categories, based on the subset
of steps mentioned above that they focus on performing effectively, and the
amount of information used to estimate the disparity for each pixel:
\begin{enumerate}
	\item \textit{Local} methods~\cite{de2011linear, hosni2013secrets,
		hosni2013fast} focus on steps (a) and (b), finding correspondences between
	pixels in the left and right images by matching simple, window-based features
	across the disparity range. Whilst fast and computationally cheap, they suffer
	in textureless/repetitive areas, and can easily estimate incorrect disparities.
	\item \textit{Global} methods~\cite{sun2003stereo, komodakis2007fast,
		besse2014pmbp, chen2014fast}, by contrast, are better suited to estimating
	accurate depths in those areas, since they enforce smoothness over disparities
	via the (possibly approximate) minimisation of an energy function defined over
	the whole image (they focus on steps (c) and (d)). However, this increased
	accuracy tends to come at a high computational cost, making these methods
	unsuitable for real-time applications.
\end{enumerate}
Semi-global matching (SGM)~\cite{hirschmuller2008stereo} bridges the gap between
local and global methods: by approximating the global methods' image-wide
smoothness constraint with the sum of several directional minimisations over the
disparity range (usually 8 or 16 directions, in a star-shaped pattern), it
produces reasonable depth in a fraction of the time.
It has thus proved highly popular in real-world systems, and many FPGA-based
approaches have been inspired by it~\cite{gehrig2009real, banz2010real,
	schmid2013stereo, honegger2014real, shan2014hardware, mattoccia2015passive}.
Other FPGA-based methods have also been presented \cite{Zhang2011,
	rahnama2017embedded, ttofis2012towards, Perri2018a, Rahnama2018}, but, whilst
typically faster than those inspired by SGM, they seldom reach the same level of
accuracy.
However, because the disparities that SGM computes for neighbouring pixels are
based on star-shaped sets of input pixels that are mostly disjoint, SGM suffers
from streaking in areas in which the data terms in some directions are weak,
whilst those in other directions are strong.
Recently, this problem has been partially addressed by an approach called More
Global Matching (MGM) \cite{facciolo2015bmvc}, which incorporates information
from two directions into each of SGM's directional minimisations; however,
because this work was not designed with FPGAs in mind, it cannot be applied
straightforwardly in an embedded context (it requires multiple passes over the
pixels in the input images, several of them in non-raster order, to compute the
bi-directional energies to be minimised).

In this paper, we present an approach inspired by MGM~\cite{facciolo2015bmvc}
that is much more amenable to real-time FPGA implementation, whilst achieving
similar accuracy and much lower power consumption. We replace the multiple
bi-directional minimisations of MGM, some of which cannot be computed in raster
order, with a single four-directional minimisation based only on pixels that are
available when processing the image as a stream. This allows us to process each
image in raster order and in a single pass, allowing us to stream data directly
from a camera connected to the FPGA and output disparity values without
requiring an intermediate buffering stage, simplifying the system
architecture and reducing latency.

This paper is structured as follows. In \autoref{sec:background}, we
review SGM~\cite{hirschmuller2008stereo} and MGM~\cite{facciolo2015bmvc}, the
algorithms that inspired our work.
In \autoref{sec:method}, we describe our method, and show how to implement it on an FPGA.
Finally, in \autoref{sec:experiments}, we evaluate our method's accuracy on the
KITTI~\cite{Geiger2012CVPR,Menze2015CVPR,Menze2015ISA} and
Middlebury~\cite{Scharstein2014} datasets, and examine its power consumption and
FPGA resource usage. By comparing it to other real-time stereo methods, we show
that it achieves a state-of-the-art balance between accuracy, speed and power
efficiency, making it desirable for use in real-time low-power systems.

\section{Background}
\label{sec:background}

\subsection{Semi-Global Matching (SGM)}
\label{sec:background-sgm}

%---
\begin{stusubfig*}{!t}
	\begin{subfigure}{.23\linewidth}
		\centering
		\includegraphics[height=\linewidth]{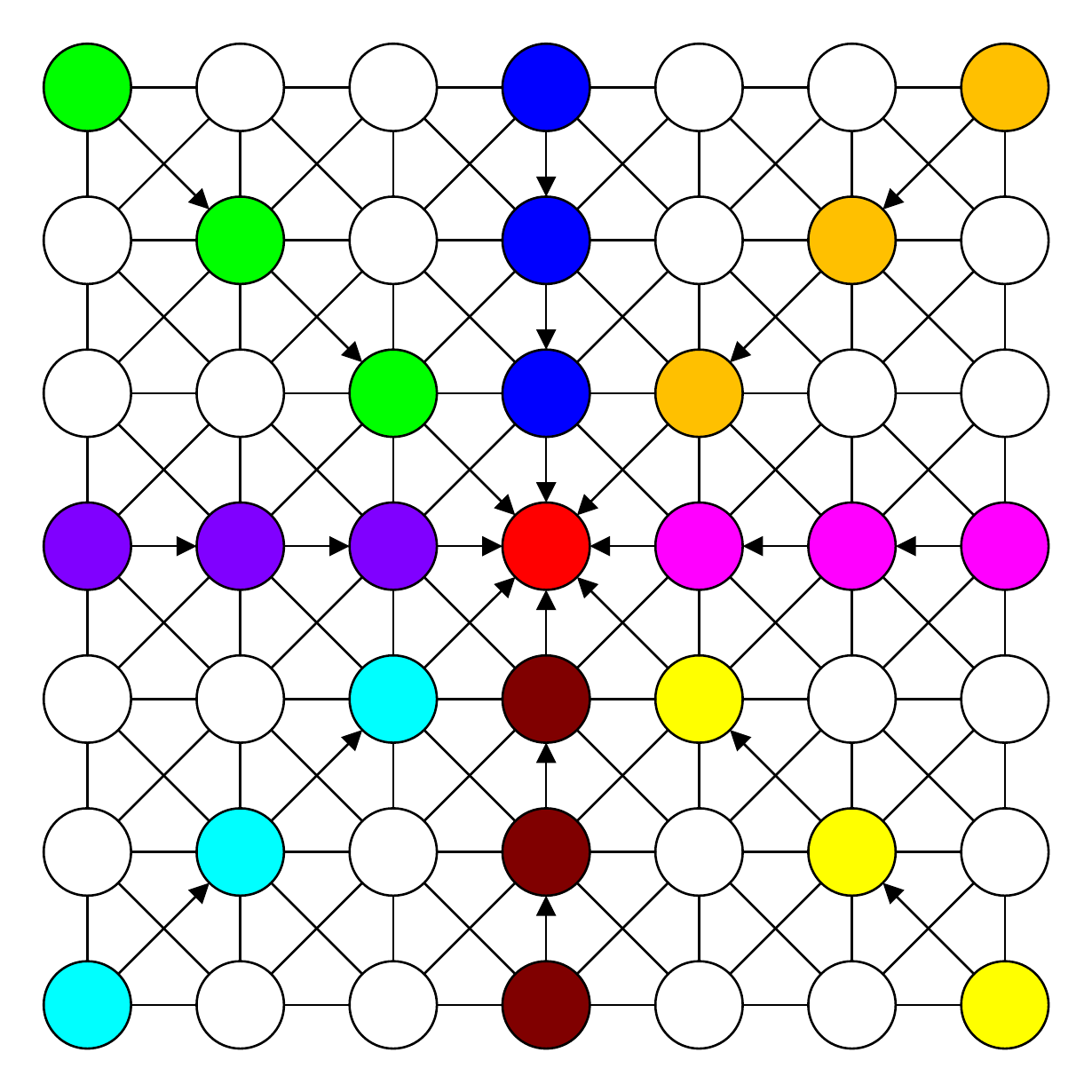}
		\caption{}
		\label{fig:sgm-full}
	\end{subfigure}%
	\hspace{4mm}%
	\begin{subfigure}{.23\linewidth}
		\centering
		\includegraphics[height=\linewidth]{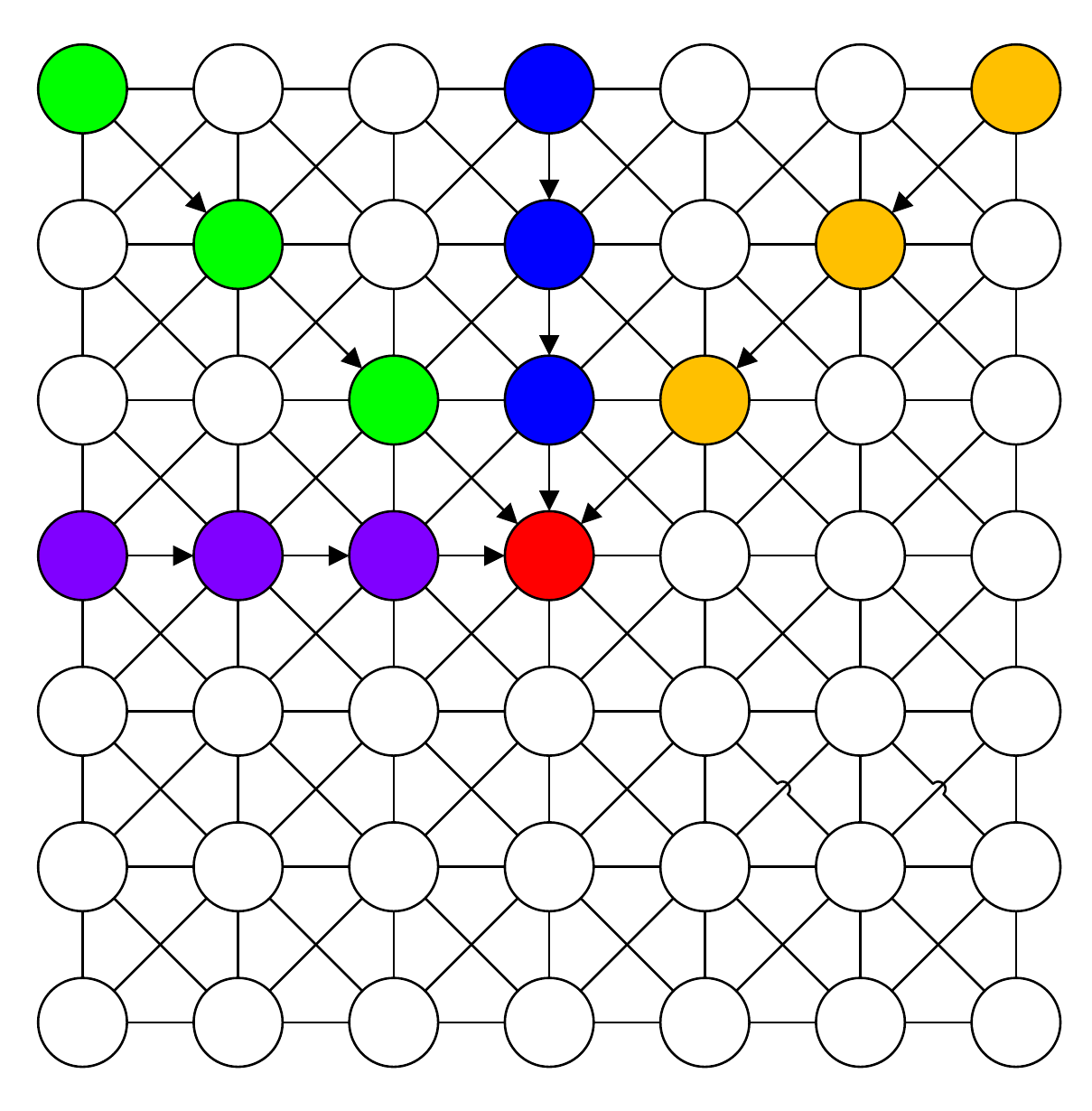}
		\caption{}
		\label{fig:sgm-raster}
	\end{subfigure}%
		\vspace{.4cm}
	\begin{subfigure}{.23\linewidth}
		\centering
		\includegraphics[height=\linewidth]{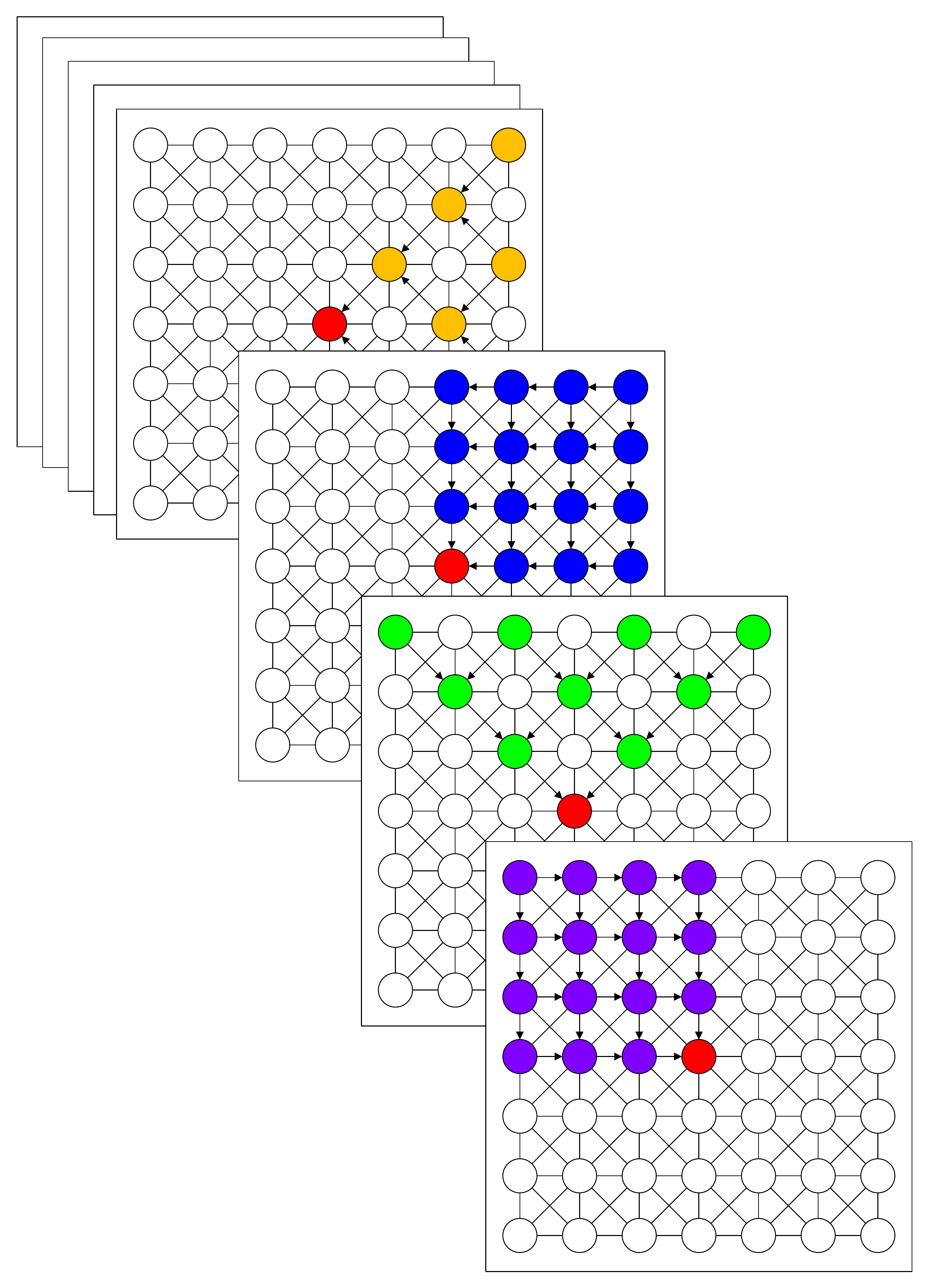}
		\caption{}
		\label{fig:mgm}
	\end{subfigure}%
	\hspace{4mm}%
	\begin{subfigure}{.23\linewidth}
		\centering
		\includegraphics[height=0.97\linewidth]{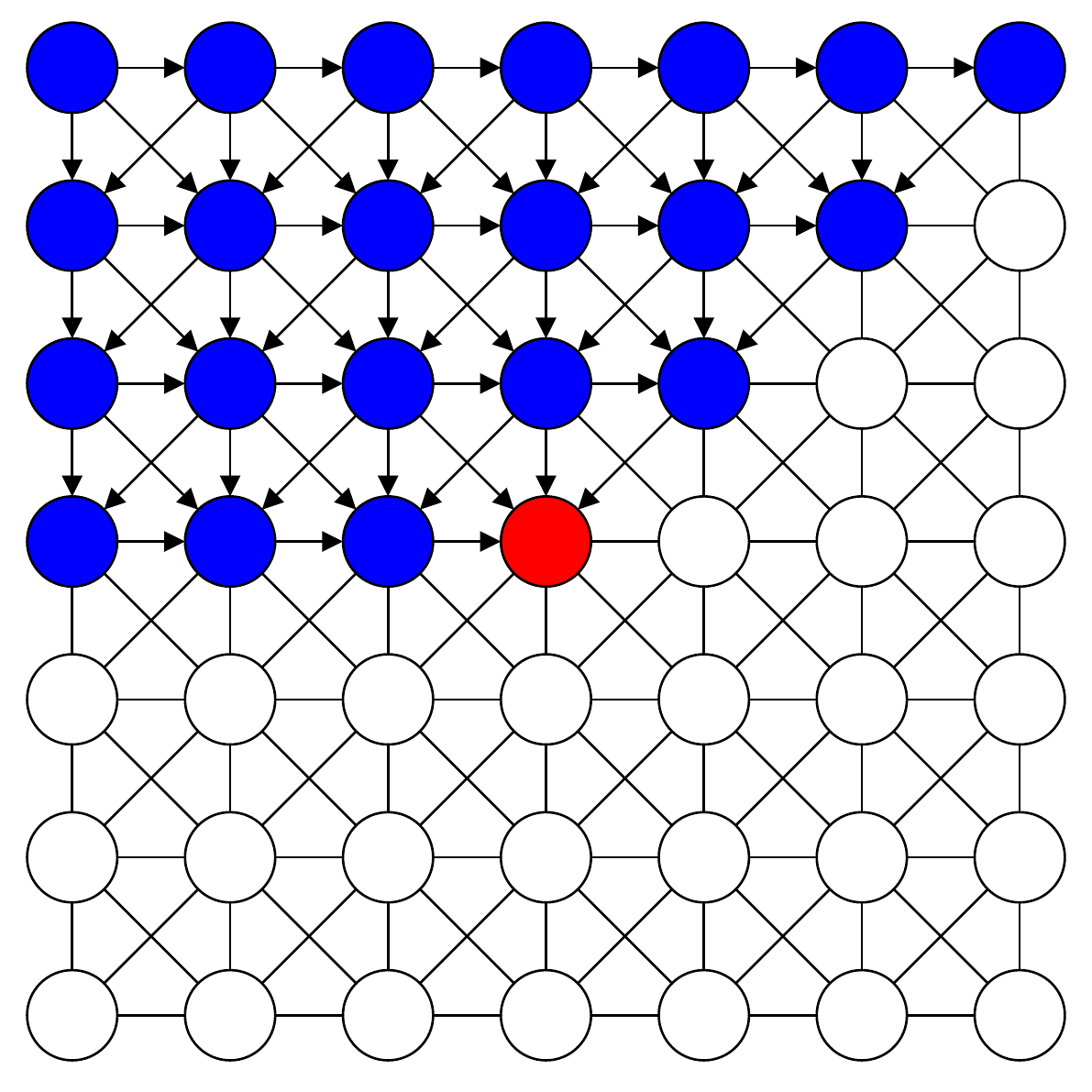}
		\caption{}
		\label{fig:ours-forward}
	\end{subfigure}%
	\vspace{-\baselineskip}
	\caption{
		The pixels used to compute $L_\mathbf{r}(\mathbf{p},\cdot)$ for pixel $\mathbf{p}$ (in red) and each scan line $\mathbf{r}$, for
		(\subref{fig:sgm-full}) a full implementation of SGM (\autoref{sec:background-sgm}), and (\subref{fig:sgm-raster}) a typical raster-based FPGA implementation of SGM (\autoref{sec:sgm-fpga}).
		Each non-red colour denotes a distinct scan line.
		In (\subref{fig:mgm}), we show the pixels that MGM would use for the same scan lines (\autoref{sec:background-mgm}).
		Finally, in (\subref{fig:ours-forward}), we show the pixels used to compute our \emph{single} cost term, allowing us to estimate disparities whilst processing pixels in a streaming fashion (note that to compute the cost vector associated with the red pixel, we require the cost vectors only from pixels that precede it in raster order).
		See \autoref{sec:method} for more details.
	}
	\label{fig:sgm}
	\vspace{-\baselineskip}
\end{stusubfig*}
%---

\noindent SGM~\cite{hirschmuller2008stereo} is a popular stereo matching method, owing to the good balance it achieves between accuracy and computational cost.
As per \cite{drory2014semi}, it aims to find a disparity map $D$ that minimises the following energy function, defined on an undirected graph $G = (I, \mathcal{E})$, with $I$ the image domain and $\mathcal{E}$ the set of edges defined by the 8-connectivity rule:
%---
\begin{equation}
\label{eq:sgm-graph}
E(D) = \sum_{\mathbf{p} \in I} C_{\mathbf{p}}(D(\mathbf{p})) + \sum_{\{\mathbf{p},\mathbf{q}\} \in \mathcal{E}} V(D(\mathbf{p}), D(\mathbf{q}))
\end{equation}
%---
Each unary term $C_{\mathbf{p}}(D(\mathbf{p}))$ denotes the `matching cost' of assigning pixel $\mathbf{p}$ in the left image the disparity $D(\mathbf{p}) \in \mathcal{D} = \left[ 0, d_{\max} \right]$. This would match it with pixel $\mathbf{p} - D(\mathbf{p}) \, \mathbf{i}$ in the right image, where $\mathbf{i} = [1,0]^\top$. 
Different matching cost functions were evaluated in \cite{Banks2001,Hirschmuller2009}.
The choice is typically based on (i) the desired invariances to nuisances (\eg changes in illumination) and (ii) computational requirements.
Each pairwise term $V(D(\mathbf{p}), D(\mathbf{q}))$ encourages smoothness by penalising disparity variations between neighbouring pixels:
%---
\begin{equation}
\label{eq:sgm-penalties}
V(d,d') = \begin{cases}
0   & \mbox{if } d = d' \\
P_1 & \mbox{if } |d - d'| = 1 \\
P_2 & \mbox{otherwise}
\end{cases}
\end{equation}
%---
The penalty $P_1$ is typically smaller than $P_2$, to avoid
over-penalising gradual disparity changes, \eg on slanted
or curved surfaces. By contrast, $P_2$ tends to be larger, so as to more
strongly discourage significant jumps in disparity.

Since the minimisation problem posed by \autoref{eq:sgm-graph} is
NP-hard, SGM approximates its solution by splitting it into several independent 1D
problems defined along scan lines. More
specifically, it associates each pixel $\mathbf{p}$ in the image with 8
scan lines, each of which follows one of the cardinal directions ($\ang{0}$,
$\ang{45}$, $\ldots$, $\ang{315}$), as per \autoref{fig:sgm-full}.
We can denote these scan lines as a vector set $R \subseteq \mathbb{R}^2$:
\begin{equation}
\begin{aligned}
R = \Bigl\{ & \scriptsize \colvec{1}{0}, \colvec{1}{1}, \colvec{0}{1}, \colvec{-1}{1}, \\
& \scriptsize \colvec{-1}{0}, \colvec{-1}{-1}, \colvec{0}{-1}, \colvec{1}{-1} \Bigr\}.
\end{aligned}
\end{equation}
Each pixel $\mathbf{p}$ is then associated with a directional cost $L_\mathbf{r}(\mathbf{p},d)$ for each direction $\mathbf{r} \in R$ and each disparity $d$.
These costs can be computed recursively via
%---
\begin{equation}
\label{eq:sgm-linecost}
\begin{aligned}
L_{\mathbf{r}}(\mathbf{p},d) = C_{\mathbf{p}}(d) \; + & \min_{d' \in \mathcal{D}} \left( L_{\mathbf{r}}(\mathbf{p} - \mathbf{r}, d') + V(d,d') \right) \\
- & \min_{d' \in \mathcal{D}} L_{\mathbf{r}}(\mathbf{p} - \mathbf{r}, d'),
\end{aligned}
\end{equation}
%---
in which $\mathbf{p} - \mathbf{r}$ refers to the pixel preceding $\mathbf{p}$ along the scan line denoted by $\mathbf{r}$.
The minimum $L_\mathbf{r}$ cost associated with $\mathbf{p} - \mathbf{r}$ is subtracted from all costs computed for $\mathbf{p}$ to prevent them growing without bound as the distance from the image edge increases \cite{hirschmuller2008stereo}.
Having computed the directional costs, SGM then sums them to form an aggregated cost volume:
%---
\begin{equation}
\label{eq:sgm-final}
L(\mathbf{p},d) = \sum_{\mathbf{r} \in R} L_{\mathbf{r}}(\mathbf{p}, d)
\end{equation}
%---
Finally, it selects each pixel's disparity using a Winner-Takes-All (WTA) approach to estimate a disparity map $D^*$:
%---
\begin{equation}
D^*(\mathbf{p}) = \argmin_{d \in \mathcal{D}} \; L(\mathbf{p},d)
\end{equation}
%---
The disparities estimated by SGM only approximate the solution of the initial
problem, for which we would need to enforce a smoothness term over the whole
image grid, but they are much less demanding to compute and, despite causing
streaking artifacts in the final disparity image, have been proven to be
accurate enough for practical purposes \cite{hirschmuller2011semi}.

One technique commonly used to filter out incorrect disparity values is to
perform an LR consistency check \cite{hirschmuller2008stereo}, which involves
computing the disparities not just of pixels in the left image, but also in the
right image, and checking that the two match (\eg that if $\mathbf{p}$ in the
left image has disparity $d$, then so does pixel $\mathbf{p} - d\mathbf{i}$ in
the right image). Observe that the disparities of pixels in the right image have
the opposite sign, \ie that assigning pixel $\mathbf{p}'$ in the right image a
disparity of $d$ matches it with pixel $\mathbf{p}' + d\mathbf{i}$ in the left
image.

Whether LR consistency checking is used or not, though, SGM has drawbacks: (i)
it suffers from streaking in textureless/repetitive regions, which the LR checks
mitigate but do not solve, (ii) there is a need to store the entire unary cost
image (or images, when checking), to allow the computation of the directional
contributions to the final cost, and (iii) there is a need for multiple passes
over the data, to recursively compute the directional components used in
\autoref{eq:sgm-final}. To deploy SGM on a limited-memory platform, \eg an FPGA,
some compromises must be made, as we now discuss.

\subsubsection{SGM on FPGAs}
\label{sec:sgm-fpga}

Some of the first implementations of SGM that were deployable on FPGA platforms
were the ones by Gehrig \etal~\cite{gehrig2009real} and Banz
\etal~\cite{banz2010real}. As the computation of the directional costs for a
pixel requires us to have already evaluated the cost function for all pixels
along the scan line (from the edge of the image), most FPGA implementations
focus only on the scan lines that would be completely available when evaluating
a pixel. If pixels in the images are available in raster order, then these will
be the three scan lines leading into the pixel from above, and the one leading
into it from its left (see \autoref{fig:sgm-raster}). Observe from
\autoref{eq:sgm-linecost} that to compute the directional costs for a pixel
$\mathbf{p}$ along a scan line $\mathbf{r}$, only its unaries $C_\mathbf{p}$ and
the cost vector $L_\mathbf{r}(\mathbf{p} - \mathbf{r}, \cdot)$ associated with
its predecessor $\mathbf{p} - \mathbf{r}$ are required.

Memory requirements are also a constraint for implementations on accelerated
platforms: when processing pixels in raster order, temporary storage is required
for the directional costs $L_\mathbf{r}$ associated with every predecessor of
the pixel being evaluated, so the more scan lines we consider, the more storage
is required from the FPGA fabric. Due to the limited resources available on
FPGAs, the choice algorithms such as \cite{gehrig2009real,banz2010real} make to
limit the number of scan lines is thus required not only to allow the processing
of pixels in raster order, but also to keep the complexity of the design low
enough to be deployed on the circuits. Despite attempts by a number of real-time
FPGA-based approaches to mitigate this
\cite{georgoulas2011real,werner2014hardware,perez2015fpga,wang2015real,cocorullo2016efficient,li2016soc,perez2016fpga},
this choice negatively impacts depth quality.

\subsection{More Global Matching (MGM) \cite{facciolo2015bmvc}}
\label{sec:background-mgm}

\noindent The streaking effect that afflicts SGM is caused by the star-shaped
pattern used when computing the directional costs (see \autoref{fig:sgm-full}):
this makes the disparity computed for each pixel depend only on a star-shaped
region of the image. To encourage neighbouring pixels to have similar
disparities, SGM relies on their `regions of influence' overlapping; however, if
the areas in which they overlap are uninformative (\eg due to limited/repetitive
texture), this effect is lost. As a result, if the contributions from some scan
lines are weak, whilst those from others are strong, the disparities of pixels
along the stronger lines will tend to be similar, whilst there may be little
correlation along the weaker scan lines: this can lead to streaking. This is an
inherent limitation of SGM, and one that is only accentuated by removing scan
lines, as is common when deploying SGM on an FPGA.

Recently, Facciolo \etal~\cite{facciolo2015bmvc} presented an extension of SGM that reduces streaking by incorporating information from multiple directions into the costs associated with each scan line.
To do this, they modify \autoref{eq:sgm-linecost} to additionally use the cost vectors of pixels on the previous scan line: see \autoref{fig:mgm}.
More specifically, when computing the cost vector $L_\mathbf{r}(\mathbf{p},\cdot)$ for pixel $\mathbf{p}$ and direction $\mathbf{r}$, they make use of the cost vector computed for the pixel $\mathbf{p} - \mathbf{r}^\bot$ ``above'' it, where ``above'' is defined relative to $\mathbf{r}$, and has the usual meaning when $\mathbf{r}$ is horizontal.
\autoref{eq:sgm-linecost} then becomes:
%---
\begin{equation}
\footnotesize
L_{\mathbf{r}}(\mathbf{p},d) = C_{\mathbf{p}}(d) + \frac{1}{2} \sum_{\mathbf{x} \in \left\{ \mathbf{r},\mathbf{r}^\bot \right\}} \min_{d' \in \mathcal{D}} \left( L_{\mathbf{r}}(\mathbf{p} - \mathbf{x}, d') + V(d,d') \right)
\end{equation}
%---
This approach was shown \cite{facciolo2015bmvc} to be more accurate than SGM, whilst running at a similar speed.
Unfortunately, the directional costs are hard to compute on accelerated platforms, and so MGM cannot easily be sped up to obtain a real-time, power-efficient algorithm.

\section{Our Approach}
\label{sec:method}

%---
\stufigstar{width=.9\linewidth}{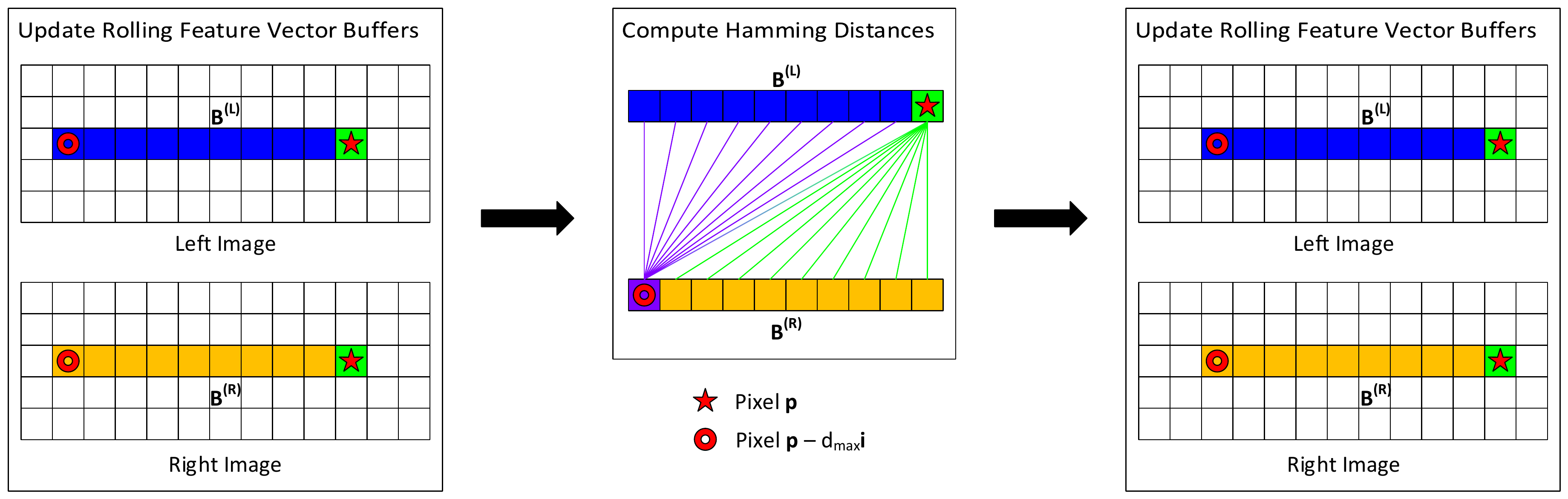}{Computing the unaries: at each pixel $\mathbf{p}$, we compute $\phi^{(\mathcal{L})}(\mathbf{p})$ and $\phi^{(\mathcal{R})}(\mathbf{p})$ and use them to update the rolling buffers $B^{(\mathcal{L})}$ and $B^{(\mathcal{R})}$. We then compute the unaries $C_\mathbf{p}^{(\mathcal{L})}(d)$ and $C_{\mathbf{p} - d_{\max}\mathbf{i}}^{(\mathcal{R})}(d)$ for all $d \in \mathcal{D}$ as the Hamming distances between the relevant feature vectors (see \autoref{eq:unaries}) before moving on to the next pixel.}{fig:unarytraversal}{!t}
%---

%---
\stufigstar{width=.9\linewidth}{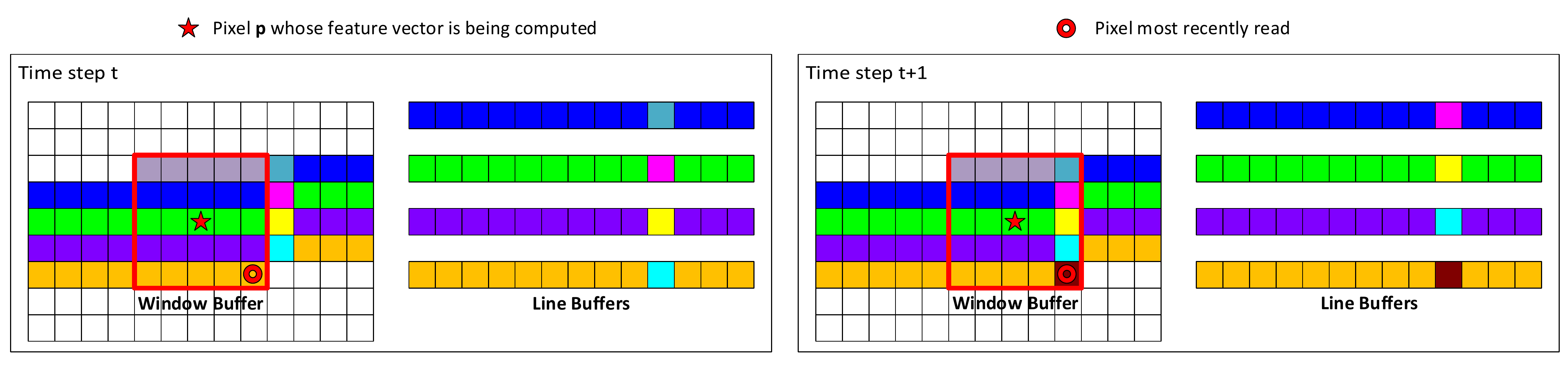}{The buffers used to compute the Census Transform feature vectors, and how they are updated (see \autoref{subsubsec:unarycomputation}).}{fig:featurevectorcomputation}{!t}
%---

\noindent MGM \cite{facciolo2015bmvc} is effective at removing streaking,
but since all but two of its directional minimisations (the purple
and green ones in \autoref{fig:mgm}, for which $\mathbf{r} = [1,0]^\top$ and
$\mathbf{r} = [1,-1]^\top$) rely on pixels that would be
unavailable when streaming the image in raster order, a \emph{full} FPGA implementation
of the algorithm is difficult to achieve (see
\autoref{sec:sgm-fpga}). One solution is to implement a cut-down
MGM that only uses those of its directional minimisations that do
work on an FPGA (\ie the purple and green ones), mirroring one way in
which SGM has been adapted for FPGA deployment \cite{banz2010real}.
However, if we limit ourselves to only one of MGM's directional minimisations
(\eg the purple one), then the `region of influence' of each pixel shrinks,
resulting in poorer disparities, and if we use both, then we are forced to use
double the amount of memory to store the cost vectors (see
\autoref{sec:sgm-fpga}).

To avoid both problems, we propose a compromise,
inspired by the way in which MGM allows each directional cost for a pixel to be
influenced by neighbouring pixels in more than one direction to mitigate
streaking. Our approach uses only a \emph{single} directional minimisation, but
one that incorporates information from \emph{all} of the directions that are
available when processing in raster order. This approach is inherently
raster-friendly, and requires a minimal amount of memory on the FPGA. When
processing the image in raster order, we compute the cost vector for each pixel
by accumulating contributions from the $4$ of its $8$ neighbours that have
already been visited and had their costs computed (the left, top-left, top and
top-right neighbours, as per \autoref{fig:ours-forward}).
Formally, if we let
\begin{equation}
\footnotesize
X = \left\{ \rightarrow,\searrow,\downarrow,\swarrow \right\} = \left\{ \colvec{1}{0}, \colvec{1}{1}, \colvec{0}{1}, \colvec{-1}{1} \right\},
\end{equation}
then we can compute the cost vector $L(\mathbf{p},\cdot)$ for pixel $\mathbf{p}$ via:
%---
\begin{equation}
\label{eq:ours}
\footnotesize
\begin{aligned}
L(\mathbf{p},d) = C_{\mathbf{p}}(d) + \frac{1}{|X|} \sum_{\mathbf{x} \in X} \biggl( & \min_{d' \in \mathcal{D}} \left( L(\mathbf{p} - \mathbf{x}, d') + V(d,d') \right) \\
- & \min_{d' \in \mathcal{D}} \left( L(\mathbf{p} - \mathbf{x}, d') \right) \biggr)
\end{aligned}
\end{equation}
Since, unlike SGM and MGM, we only use a single minimisation, this is like \autoref{eq:sgm-final} in those approaches, letting us obtain each pixel's cost vector in a single pass over the image.

In our implementation, we use the Census Transform (CT) \cite{Zabih1994} to
compute the unaries $C_\mathbf{p}$.
CT is robust to illumination changes between the images, and can be computed efficiently and in a raster-friendly way (see \autoref{subsubsec:unarycomputation}).
Moreover, the Hamming distance between two CT feature vectors can be computed efficiently, and provides a good measure of their similarity.
We compute the pixel costs $L(\mathbf{p}, d)$ simultaneously for both the left
and right images, thanks to our FPGA implementation (see
\autoref{subsubsec:directionalcomputation}). After selecting the best disparity
in each image with a WTA approach, we process the disparities with a median
filter (in a raster-friendly way) to reduce noise in the output.
Finally, we validate our disparities with a standard LR check to discard
inconsistent results \cite{hirschmuller2008stereo}, using a threshold of $1$
disparity or $3\%$, whichever is greater.

\subsection{FPGA Implementation}

%---
\stufigstar{width=.9\linewidth}{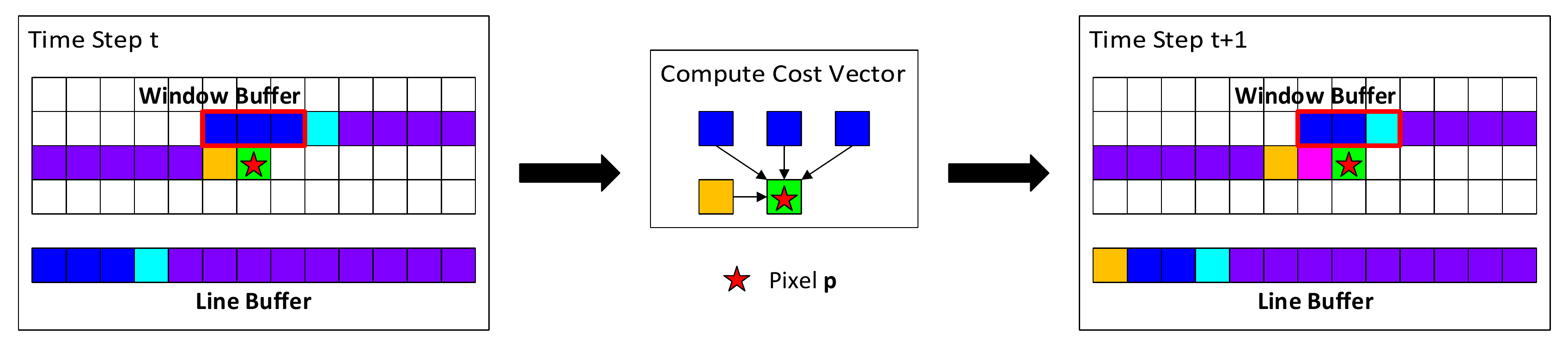}{The buffers used to compute the cost vectors, and how they are updated (see \autoref{subsubsec:directionalcomputation}).}{fig:directionalcomputation}{!t}
%---

\noindent Having described our approach conceptually, we now show how to
implement it on an FPGA. By contrast to the previous sections, in which we only
showed how to compute the disparities for pixels in the left image, here we
describe how to compute the disparities for the pixels in both images
efficiently to support LR consistency checking (see
\autoref{sec:background-sgm}). Notationally, we distinguish between the unary
costs and cost vectors for the two images using the superscripts $(\mathcal{L})$
and $(\mathcal{R})$.

Two main steps are involved: (i) the computation of the unary costs
$C_\mathbf{p}^{(\mathcal{L})}(\cdot)$ and $C_\mathbf{p}^{(\mathcal{R})}(\cdot)$,
and (ii) the recursive computation of the cost vectors
$L^{(\mathcal{L})}(\mathbf{p},\cdot)$ and $L^{(\mathcal{R})}(\mathbf{p},\cdot)$.
Implementing these steps efficiently on an FPGA requires understanding the
hardware, which essentially consists of a set of programmable logic blocks that
can be wired together in different ways and independently programmed to perform
different functions. This architecture naturally lends itself to an efficient
pipeline-processing style, in which data is fed into the FPGA in a stream and
different logic blocks try to operate on different pieces of data concurrently
in each clock cycle. In practice, the steps we consider here all involve
processing images, with the data associated with the images' pixels being
streamed into the FPGA in raster order, as we now describe.

\subsubsection{Unary Computation}
\label{subsubsec:unarycomputation}

Each unary $C_\mathbf{p}^{(\mathcal{L})}(d)$, which denotes the cost of
assigning pixel $\mathbf{p}$ in the left image a disparity of $d$, is computed
as the Hamming distance $\mathcal{H}$ between the feature vector
$\phi^{(\mathcal{L})}(\mathbf{p})$ of pixel $\mathbf{p}$ in the left image and
the feature vector $\phi^{(\mathcal{R})}(\mathbf{p} - d\mathbf{i})$ of pixel
$\mathbf{p} - d\mathbf{i}$ in the right image.
($\phi^{(\mathcal{L})}(\mathbf{p})$ is computed by applying the Census Transform
\cite{Zabih1994} to a $W \times W$ window around $\mathbf{p}$ in the left image,
and analogously for $\phi^{(\mathcal{R})}$.) Conversely,
$C_\mathbf{p}^{(\mathcal{R})}(d)$ becomes the Hamming distance between
$\phi^{(\mathcal{R})}(\mathbf{p})$ and $\phi^{(\mathcal{L})}(\mathbf{p} +
d\mathbf{i})$.

As per \autoref{fig:unarytraversal}, we traverse both the left and right images
simultaneously in raster order, computing $\phi^{(\mathcal{L})}(\mathbf{p})$ and
$\phi^{(\mathcal{R})}(\mathbf{p})$ for each pixel $\mathbf{p}$ as we go, and
maintaining rolling buffers $B^{(\mathcal{L})}$ and $B^{(\mathcal{R})}$ of
feature vectors for the most recent $d_{\max} + 1$ pixels in each image, \ie
$B^{(\mathcal{L})} = [\phi^{(\mathcal{L})}(\mathbf{p} - d\mathbf{i}) : d \in
\mathcal{D}]$, and analogously for the right image. After computing the feature
vectors for pixel $\mathbf{p}$, we compute unaries for all $d \in \mathcal{D}$:
\begin{equation}
\label{eq:unaries}
\scriptsize
\begin{aligned}
C_\mathbf{p}^{(\mathcal{L})}(d) &= \mathcal{H}(\phi^{(\mathcal{L})}(\mathbf{p}), \phi^{(\mathcal{R})}(\mathbf{p} - d\mathbf{i})) \\
C_{\mathbf{p} - d_{\max}\mathbf{i}}^{(\mathcal{R})}(d) &= \mathcal{H}(\phi^{(\mathcal{L})}(\mathbf{p} + (d - d_{\max})\mathbf{i}), \phi^{(\mathcal{R})}(\mathbf{p} - d_{\max}\mathbf{i}))
\end{aligned}
\end{equation}
Note that we compute the unaries for right image pixels just before they leave
$B^{(\mathcal{R})}$, since it is only at that point that we have accumulated the
feature vectors for all of the relevant left image pixels in $B^{(\mathcal{L})}$
(see \autoref{fig:unarytraversal}).

In practice, to efficiently compute the feature vectors, we must maintain a $W
\times W$ window of the pixels surrounding $\mathbf{p}$ that can be used to
compute the Census Transform \cite{Zabih1994}. As shown in
\autoref{fig:featurevectorcomputation}, we store this in a window buffer (local
registers on an FPGA that can be used to store data to which we require
instantaneous access). To keep the window buffer full, we must read ahead of
$\mathbf{p}$ by slightly over $\lfloor W / 2 \rfloor$ rows. Separately, we
maintain pixels from the rows above/below $\mathbf{p}$ in line buffers (regions
of memory on an FPGA that can store larger amounts of data but can only provide
a single value per clock cycle). As shown in
\autoref{fig:featurevectorcomputation}, some pixels are in both the window
buffer and one of the line buffers. When moving from one pixel to the next, we
update the window buffer and line buffers as shown in
\autoref{fig:featurevectorcomputation}. Notice how the individually marked
pixels in the line buffers are shifted upwards to make way for the new, brown
pixel that is being read in (to both the orange line buffer and the window
buffer), and how the turquoise pixel is removed from the blue line buffer but
added to the top-right of the window buffer. All of these operations can be
implemented very efficiently on an FPGA.

\subsubsection{Cost Vector Computation}
\label{subsubsec:directionalcomputation}

Once the unaries have been computed, the next step is to compute the
$L(\mathbf{p},d)$ values (\ie the cost vector) for each pixel using
\autoref{eq:ours}. This again involves a walk over the image domain in raster
order. In this case, computing the cost vector for each pixel $\mathbf{p}$ uses
the cost vectors of the pixels $\mathbf{p} - \mathbf{x}$, for each $\mathbf{x}
\in X$ (\ie the $3$ pixels above $\mathbf{p}$ and the pixel to its left). As a
result, these must be in memory when the cost vector for $\mathbf{p}$ is
computed.

In practice, as shown in \autoref{fig:directionalcomputation}, we divide the
relevant cost vectors between several different locations in memory: (i) a line
buffer whose size is equal to the width of the image, (ii) a window buffer that
holds the cost vectors for the $3$ pixels above $\mathbf{p}$, and (iii) a
register that holds the cost vector for the pixel to its left (the yellow pixel
in \autoref{fig:directionalcomputation}). This provides us with instantaneous
access to the cost vectors that we need to compute the cost vector for
$\mathbf{p}$, whilst keeping track of the cost vectors for the pixels that we
will need to compute the cost vectors for upcoming pixels (via the line buffer).
When moving from one pixel to the next, we update the window buffer and line
buffer as shown in \autoref{fig:directionalcomputation}. For the actual
computation of $L(\mathbf{p},d)$, we rewrite \autoref{eq:ours} as follows:
\begin{equation}
\label{eq:ours-optimised}
\scriptsize
\begin{aligned}
L(\mathbf{p},d) = & \; C_{\mathbf{p}}(d) + \frac{1}{|X|} \sum_{\mathbf{x} \in X} \biggl( \\
& \; \begin{aligned} \min{} \Bigl\{ & L(\mathbf{p} - \mathbf{x},d), L(\mathbf{p} - \mathbf{x},d-1) + P_1, \\
& L(\mathbf{p} - \mathbf{x},d+1) + P_1, \min_{d' \in \mathcal{D}} (L(\mathbf{p} - \mathbf{x},d') + P_2) \Bigr\} \\
\end{aligned} \\
- & \min_{d' \in \mathcal{D}} \left( L(\mathbf{p} - \mathbf{x}, d') \right) \biggr)
\end{aligned}
\end{equation}
This allows for a more optimal implementation in which we store $\min_{d' \in \mathcal{D}} (L(\mathbf{p} - \mathbf{x},d')$ to avoid repeat computations.

%---
\begin{table*}[!t]
	\centering
	\footnotesize
	\begin{tabular}{ccccccc}
		\toprule
		\textbf{Method} & \textbf{D1 Valid} & \textbf{Density} & \textbf{D1 Interpolated} & \textbf{Runtime} & \textbf{Environment} & \textbf{Power Consumption (W)} \\
		&&&&&& (approx.) \\
		\midrule
    Ours & 4.8\% & 85.0\% & 9.9\% & 0.014s & FPGA (Xilinx ZC706) & 3 \\
		\midrule
		DeepCostAggr \cite{kuzmin2017} & -- & 99.98\% & 6.3\% & 0.03s & Nvidia GTX Titan X & 250 \\
		CSCT+SGM+MF \cite{hernandez2016embedded} & -- & 100\% & 8.2\% & 0.006s & Nvidia GTX Titan X & 250 \\
		\bottomrule
	\end{tabular}
	\caption{The quantitative results of our approach, in comparison to state-of-the-art GPU-based real-time methods, on the Stereo 2015 subset of the KITTI dataset \cite{Menze2015CVPR,Menze2015ISA}. D1 Valid: error rate on the pixels surviving the LR check; Density: \% of pixels output by the algorithm (in our case after the LR check); D1 Interpolated: error rate after interpolating according to the KITTI protocol. We use a threshold of $3$ disparity values or 5\%, whichever is greater (\ie the standard thresholds for KITTI).
		Our approach is able to produce post-interpolation results that are within striking distance of existing methods, whilst being two orders of magnitude more power-efficient and requiring much less computational power.
	}
	\label{tbl:quantitativeresults-kitti}
\end{table*}
%---

%---
\begin{stusubfig*}{!t}
	\begin{subfigure}{.24\linewidth}
		\centering
		\includegraphics[height=.3\linewidth]{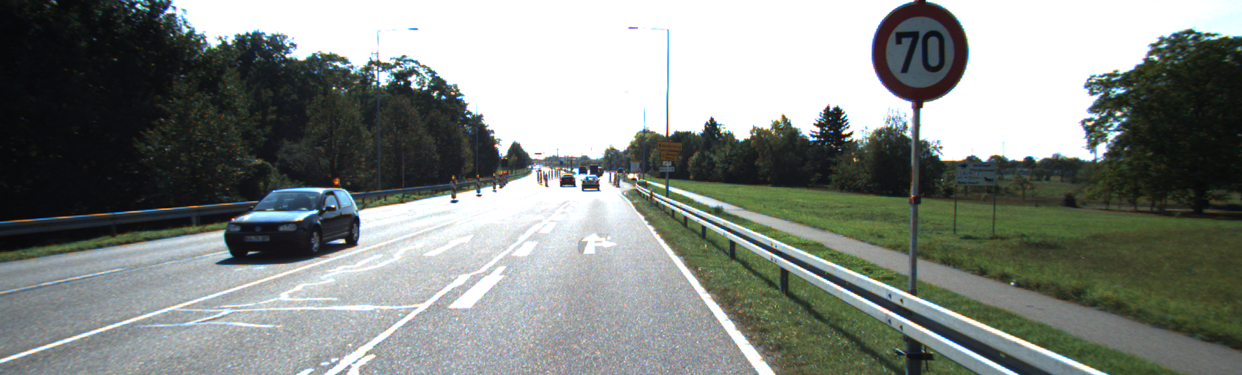}
	\end{subfigure}%
	\hspace{1px}
	\begin{subfigure}{.24\linewidth}
		\centering
		\includegraphics[height=.3\linewidth]{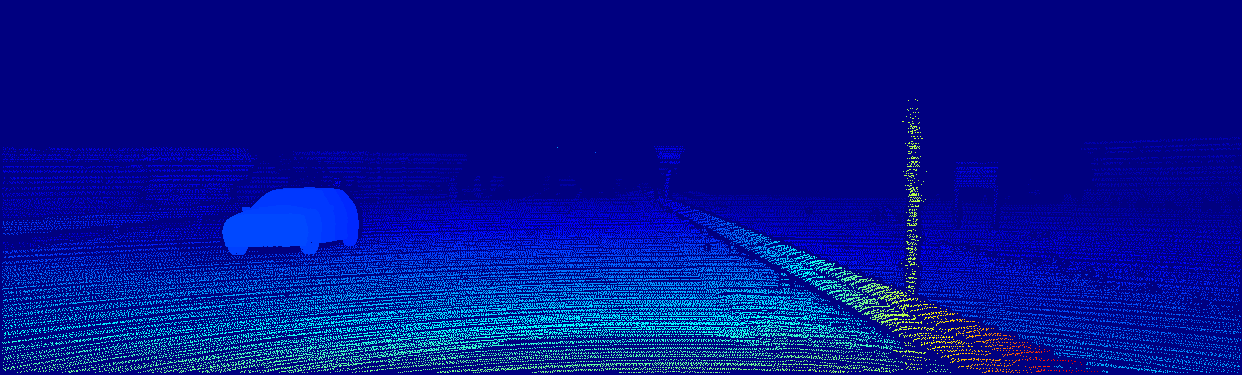}
	\end{subfigure}%
	\hspace{1px}
	\begin{subfigure}{.24\linewidth}
		\centering
		\includegraphics[height=.3\linewidth]{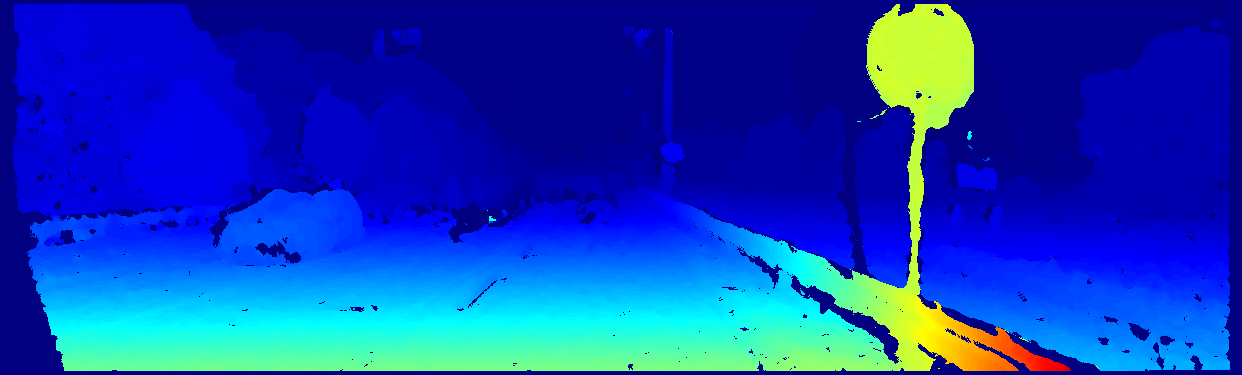}
	\end{subfigure}%
	\hspace{1px}
	\begin{subfigure}{.24\linewidth}
		\centering
		\includegraphics[height=.3\linewidth]{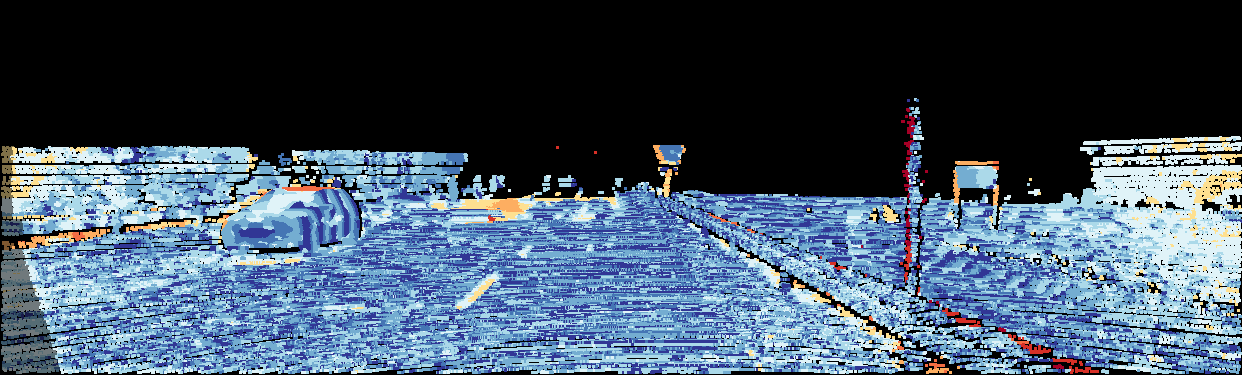}
	\end{subfigure}%
	\\[0.2cm]
	\begin{subfigure}{.24\linewidth}
		\centering
		\includegraphics[height=.3\linewidth]{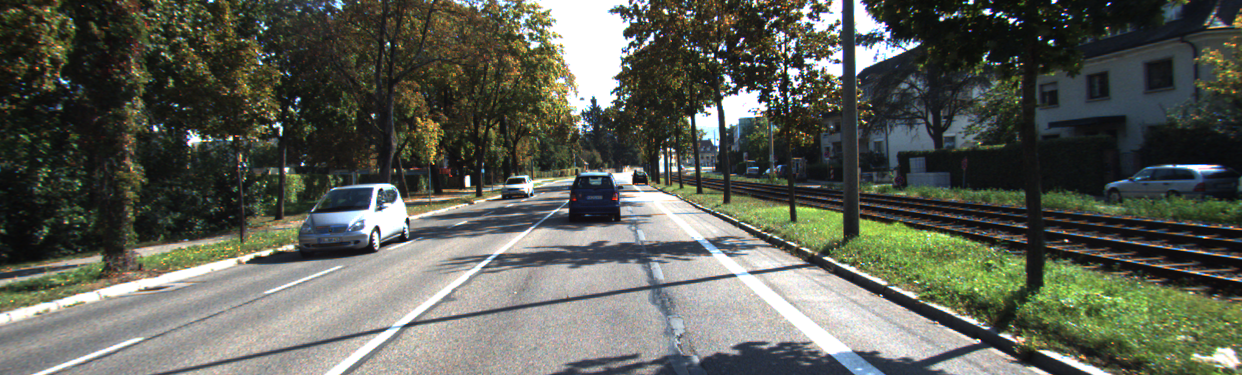}
	\end{subfigure}%
	\hspace{1px}
	\begin{subfigure}{.24\linewidth}
		\centering
		\includegraphics[height=.3\linewidth]{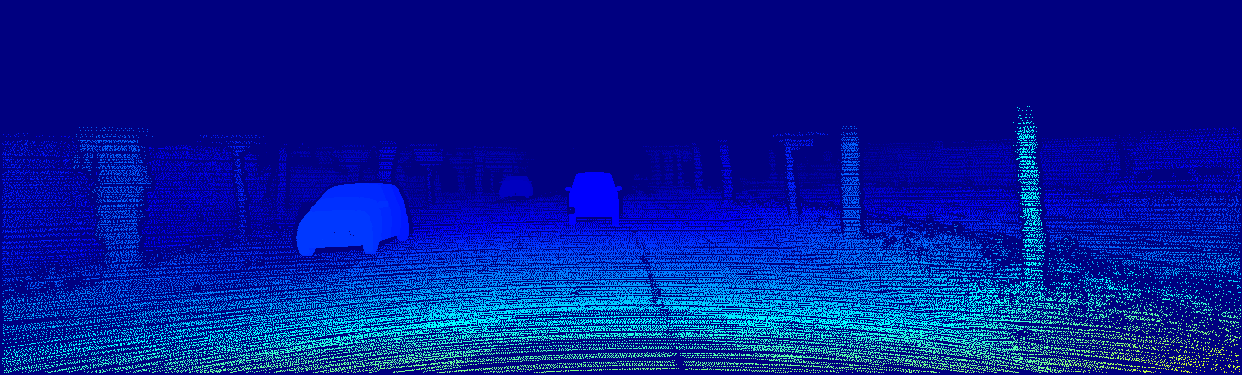}
	\end{subfigure}%
	\hspace{1px}
	\begin{subfigure}{.24\linewidth}
		\centering
		\includegraphics[height=.3\linewidth]{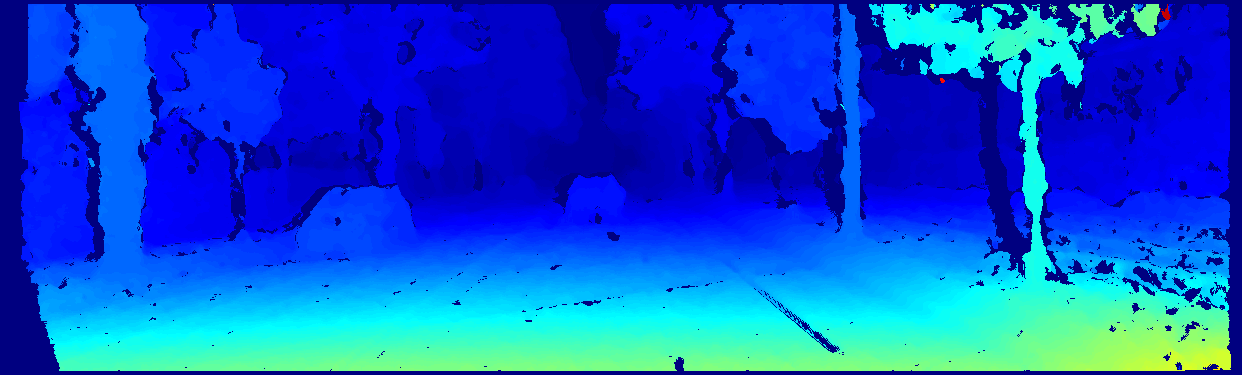}
	\end{subfigure}%
	\hspace{1px}
	\begin{subfigure}{.24\linewidth}
		\centering
		\includegraphics[height=.3\linewidth]{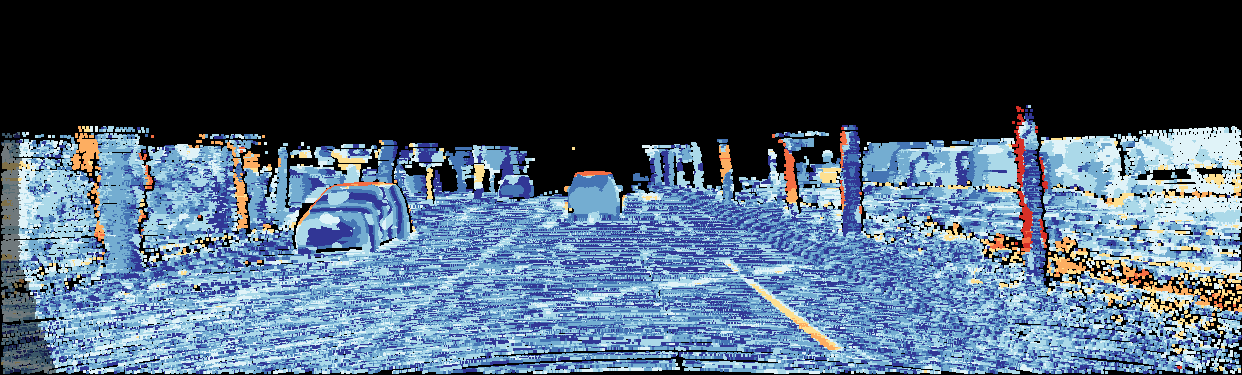}
	\end{subfigure}%
	\\[0.2cm]
	\begin{subfigure}{.24\linewidth}
		\centering
		\includegraphics[height=.3\linewidth]{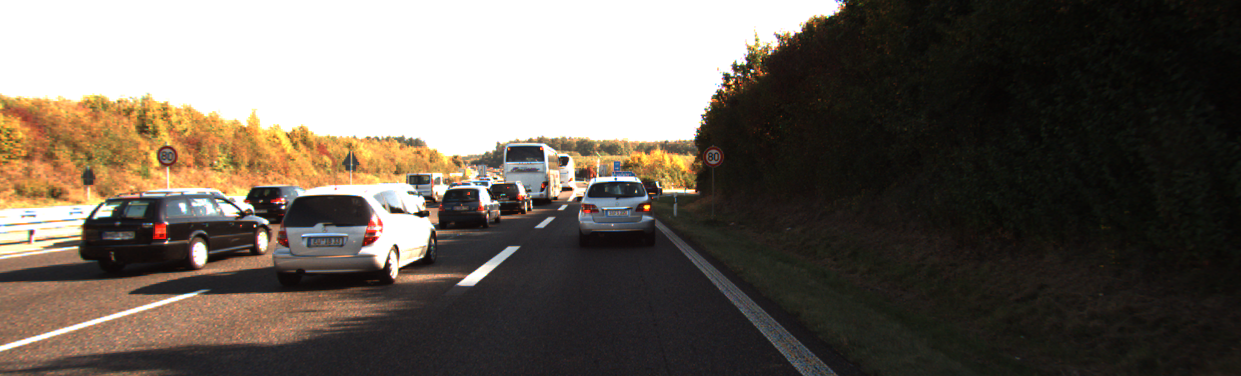}
	\end{subfigure}%
	\hspace{1px}
	\begin{subfigure}{.24\linewidth}
		\centering
		\includegraphics[height=.3\linewidth]{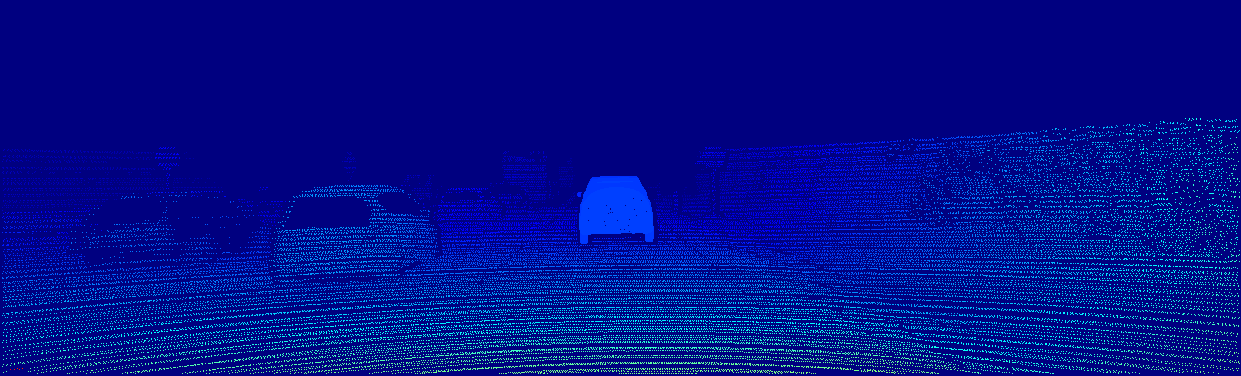}
	\end{subfigure}%
	\hspace{1px}
	\begin{subfigure}{.24\linewidth}
		\centering
		\includegraphics[height=.3\linewidth]{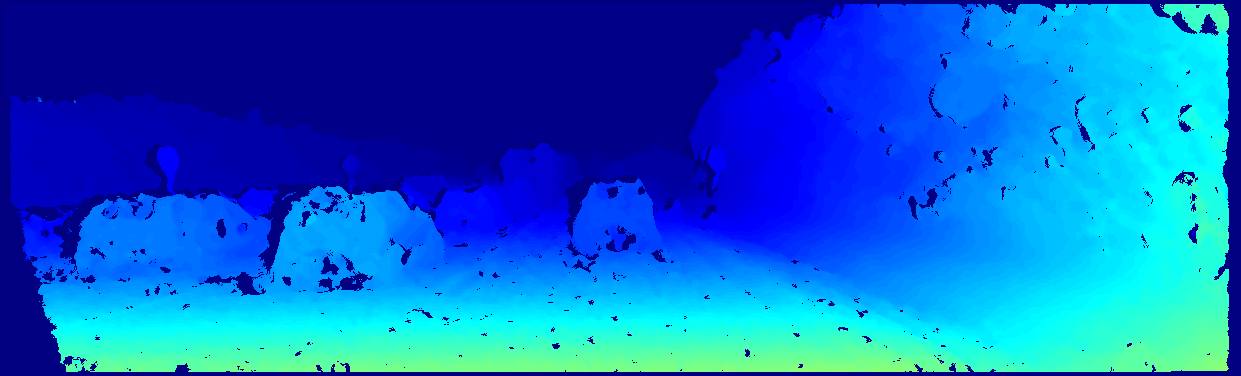}
	\end{subfigure}%
	\hspace{1px}
	\begin{subfigure}{.24\linewidth}
		\centering
		\includegraphics[height=.3\linewidth]{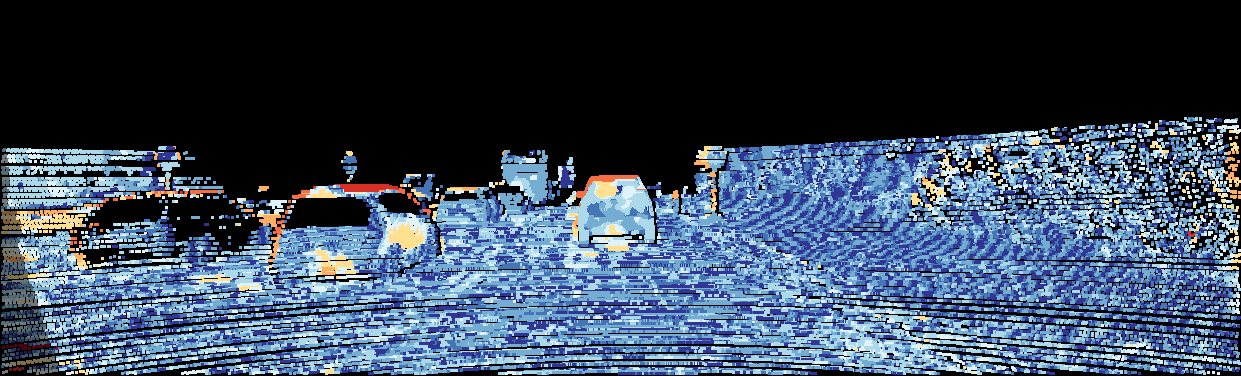}
	\end{subfigure}%
	\\[0.2cm]
	\begin{subfigure}{.24\linewidth}
		\centering
		\includegraphics[height=.3\linewidth]{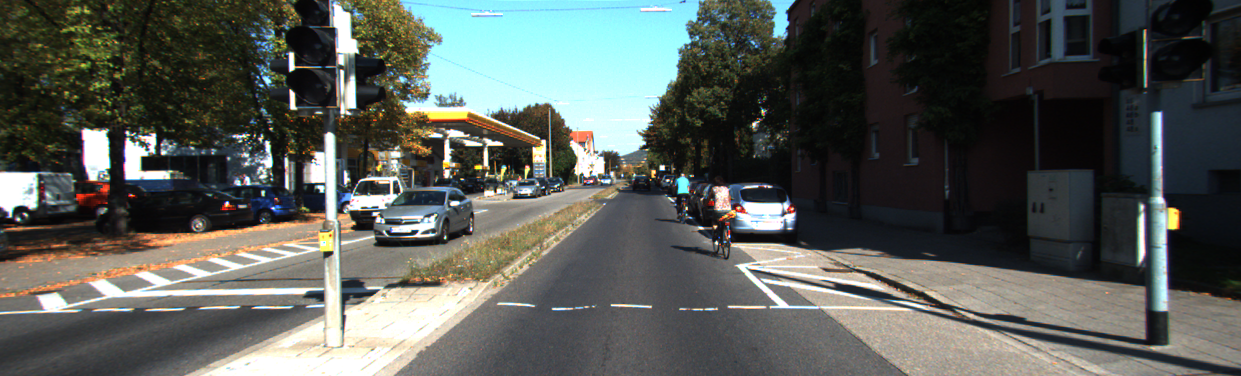}
	\end{subfigure}%
	\hspace{1px}
	\begin{subfigure}{.24\linewidth}
		\centering
		\includegraphics[height=.3\linewidth]{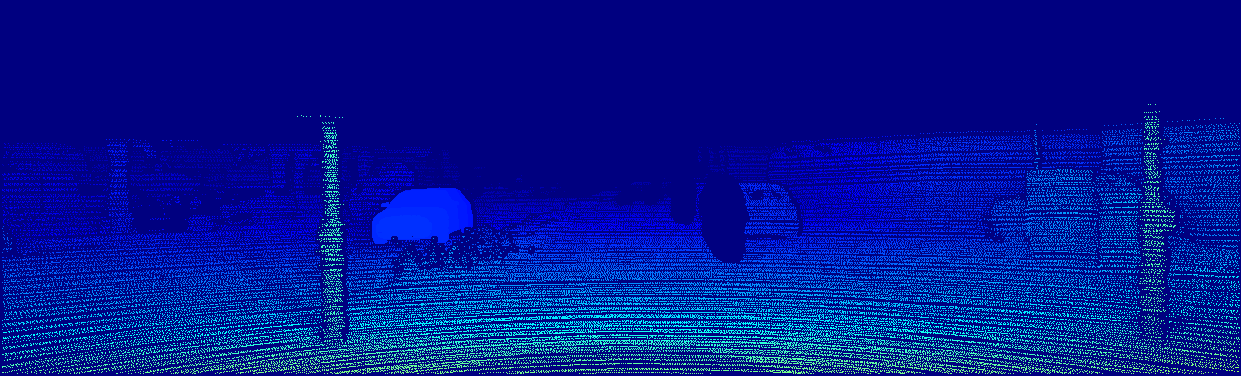}
	\end{subfigure}%
	\hspace{1px}
	\begin{subfigure}{.24\linewidth}
		\centering
		\includegraphics[height=.3\linewidth]{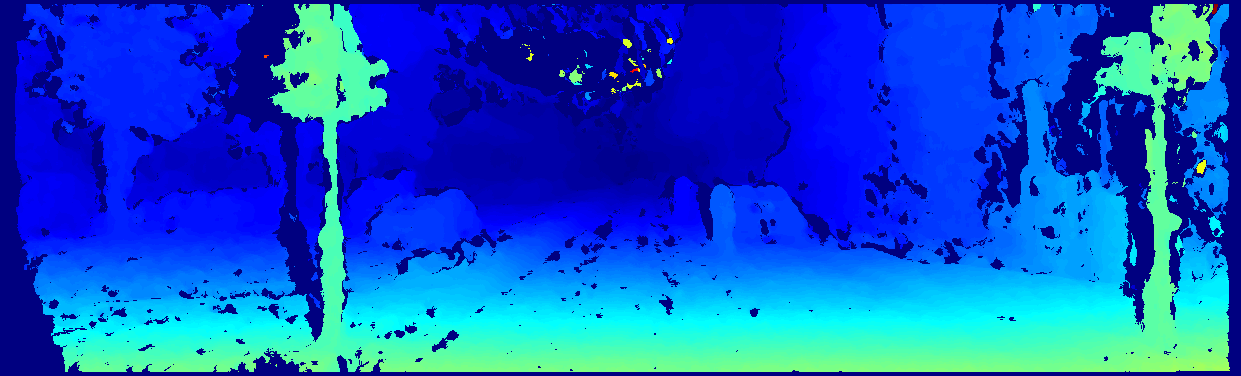}
	\end{subfigure}%
	\hspace{1px}
	\begin{subfigure}{.24\linewidth}
		\centering
		\includegraphics[height=.3\linewidth]{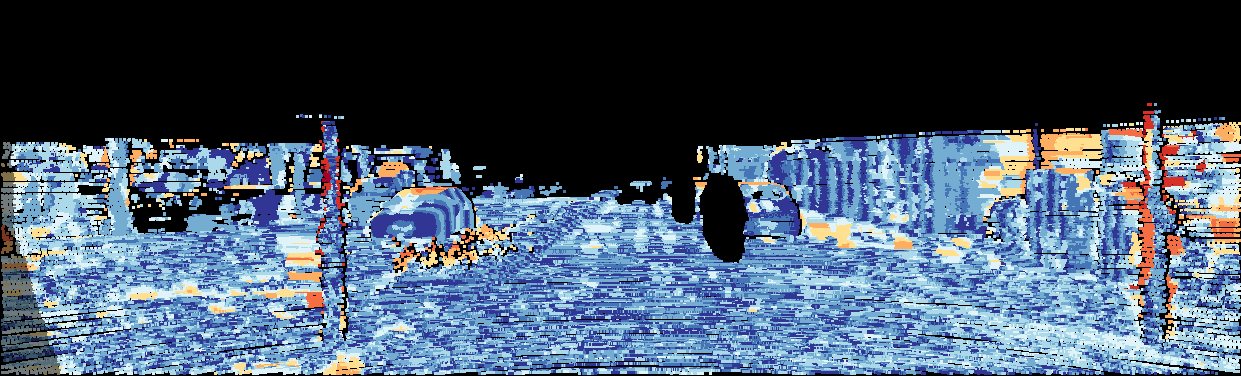}
	\end{subfigure}%
	\caption{Qualitative examples of our approach on KITTI frames \cite{Menze2015CVPR,Menze2015ISA}. Column 1: input left eye image; column 2: ground truth disparities; column 3: the disparities produced by our approach; column 4: error image (blue = low, red = high).}
	\label{fig:qualitativeresults-kitti}
\end{stusubfig*}
%---

%---
\begin{table*}[!t]
	\centering
	\footnotesize
	\begin{tabular}{ccccccccccccc}
		\toprule
		\multirow{2}{*}{\textbf{Method}} & \multicolumn{3}{c}{\textbf{Cones}} & \multicolumn{3}{c}{\textbf{Teddy}} & \multicolumn{3}{c}{\textbf{Tsukuba}} & \multicolumn{3}{c}{\textbf{Venus}} \\
		& non-occ. & all & disc. & non-occ. & all & disc. & non-occ. & all & disc. & non-occ. & all & disc. \\
		\midrule
		Ours & \textbf{3.4} & \textbf{8.9} & 10.3 & 8.2 & 14.6 & 22.4 & 9.7 & 11.2 & 31.2 & 1.0 & 1.6 & 11.9 \\
		\midrule
		\cite{banz2010real}, 4 paths &  9.5 & -- & -- & 13.3 & -- & -- &  6.8 & -- & -- &  4.1 & -- & -- \\ % non occ
		\cite{banz2010real}, 8 paths &  8.4 & -- & -- & 11.4 & -- & -- &  4.1 & -- & -- &  2.7 & -- & -- \\ % still non occ, do we care about 8 paths?
		\cite{gehrig2009real}        & -- &  9.5 & -- & -- & 13.3 & -- & -- & 5.9 & -- & -- & 3.9 & -- \\ % all pixels
		\cite{shan2014hardware}      & 3.5 & 11.1 & \textbf{9.6} & 7.5 & 14.7 & 19.4 & \textbf{3.6} & \textbf{4.2} & 14.0 & \textbf{0.5} & \textbf{0.9} & \textbf{2.8} \\
		\cite{ttofis2012towards}     & 17.1 & 25.9 & 25.8 & 21.5 & 28.1 & 28.8 & 4.5 & 6.0 & \textbf{12.7} & 6.0 & 7.5 & 18.2 \\
		\cite{Zhang2011} & 5.4 & 11.0 & 13.9 & 7.2 & 12.6 & \textbf{17.4} & 3.8 & 4.3 & 14.2 & 1.2 & 1.7 & 5.6 \\
		\cite{Perri2018a}& 9.3 & 11.1 & 17.5 & \textbf{6.0} & \textbf{7.4} & 18.7 & 8.8 & 16.4 & 20.0 & 3.9 & 12.0 & 10.3 \\
		\bottomrule
	\end{tabular}
	\caption{The accuracy of some FPGA-based methods on Middlebury images \cite{scharstein2002taxonomy,scharstein2003high}.
		The \%'s are of pixels with a disparity error $> 1$ pixel from ground truth: non-occ.\ = non-occluded pixels only, disc.\ = pixels near discontinuities only.
	}
	\label{tbl:quantitativeresults-middlebury}
	\vspace{-\baselineskip}
\end{table*}
%---

\section{Experiments}
\label{sec:experiments}

\noindent We evaluate our approach on the KITTI
\cite{Menze2015CVPR,Menze2015ISA} and Middlebury
\cite{scharstein2002taxonomy,scharstein2003high} datasets. We then compare our
frame rate across different resolutions and disparity ranges vs.\ competing
FPGA-based approaches. Finally, we break down the FPGA resource costs,
associated power consumption, as well as accuracy of our approach for several
variations of our design.

On KITTI, we compare our approach to the only two published approaches from the
benchmark that are able to achieve state-of-the-art performance in real time
\cite{kuzmin2017,hernandez2016embedded}, both of which require a powerful GPU
(an Nvidia GTX Titan X) to run. Since, unlike these approaches, our approach
does not naturally produce disparities for every single pixel in the image, we
interpolate as specified by the KITTI evaluation protocol in order to make our
results comparable with theirs. As shown in
\autoref{tbl:quantitativeresults-kitti}, we are able to achieve
post-interpolation results that are competitive on accuracy with these
approaches, whilst significantly reducing the power consumption and the compute
power required. Furthermore, compared to the additional power efficient
implementation reported in \cite{hernandez2016embedded} (on a Nvidia Tegra X1),
which achieves 13.8 fps with 10 Watts when scaled to the KITTI dataset, our
system is more than five times faster, whilst consuming less than a third of the
power. Moreover, we are able to achieve an even better error rate (4.8\%)
pre-interpolation, with a density of 85\%. For some applications, this may in
practice be more useful than having poorer disparities over the whole image.
Qualitative examples of our approach on KITTI frames are shown in
\autoref{fig:qualitativeresults-kitti}.

On Middlebury, we show in \autoref{tbl:quantitativeresults-middlebury} and
\autoref{tbl:quantitativeresults-performance} that we are able to achieve
comparable accuracy to a number of other FPGA-based methods, whilst either
running at a much higher frame-rate (c.f.\
\cite{gehrig2009real,ttofis2012towards}), using simpler, cheaper hardware (c.f.\
\cite{shan2014hardware}) or handling greater disparity ranges (c.f.\
\cite{Perri2018a, Zhang2011}). \autoref{fig:qualitativeresults-middlebury} shows
some examples of our approach on the four most commonly used Middlebury images,
namely \emph{Cones}, \emph{Teddy}, \emph{Tsukuba} and \emph{Venus}.

%---
\begin{table}[!t]
	\centering
	\scriptsize
	\begin{tabular}{ccccc}
		\toprule
		\textbf{Method} & \textbf{Resolution} & \textbf{Disparities} & \textbf{FPS} & \textbf{Environment} \\ % $d_{\max} + 1$
		\midrule
		\multirow{4}{*}{Ours} &  384x288 &  32 & 301 & \multirow{4}{*}{Xilinx ZC706} \\ %tsukuba
		&  450x375 &  64 & 198 &                                      \\ %teddy, cones
		&  640x480 & 128 & 109 &                                      \\ % vga
		& 1242x375 & 128 & 72  &                                      \\ % kitti
		\midrule
		\multirow{2}{*}{\cite{banz2010real}, 4 paths} & 640x480 &  64 & 66--167 & \multirow{2}{*}{Xilinx Virtex 5} \\ %  LX220T-1
		& 640x480 & 128 & 37--103 & \\
		\midrule
		\cite{gehrig2009real}        & 340x200 & 64 &  27 & Xilinx Virtex 4 \\ % all pixels - FX140
		\midrule
		\cite{mattoccia2015passive}  & 640x480 & 32 &  $\ge 30$ & Xilinx Spartan 6 LX \\ % all pixels
		\midrule
		\multirow{4}{*}{\cite{shan2014hardware}}      &   352x288 &  64 & 1121 & \multirow{4}{*}{Altera Stratix IV} \\ % all pixels - EP4SGX230
		&   640x480 &  64 &  357 & \\
		&  1024x768 & 128 &  129 & \\
		& 1920x1080 & 256 & 47.6 & \\
		\midrule
		\multirow{4}{*}{\cite{ttofis2012towards}} & 320x240 &  64 &  115 & \multirow{4}{*}{Xilinx Virtex 5} \\ % all pixels, not SGM-based - LX110T
		& 320x240 & 128 &   66 & \\
		& 640x480 &  64 &   30 & \\
		& 800x600 &  64 &   19 & \\
		\midrule
		\cite{schmid2013stereo}      & 1024x508 & 128 &  15 & Xilinx Spartan 6 \\
		\midrule
		\cite{honegger2014real}      & 752x480 & 32 &  $\ge 60$ & Xilinx Artix 7 \\ % XC7A100T
		\midrule
		\cite{Zhang2011} & 1024x768 & 64 & 60 & Altera EP3SL150 \\
		\midrule
		\cite{Perri2018a} & 640x480 & 32 & 101 & Xilinx Zynq-7000 \\ % XC7Z045
		\bottomrule
	\end{tabular}
	\caption{The frame rates we can achieve, in comparison to those achieved by other FPGA-based methods, for multiple resolution/disparity range combinations.}
	\label{tbl:quantitativeresults-performance}
	\vspace{-\baselineskip}
\end{table}
%---

%---
\begin{stusubfig}{!t}
	\begin{subfigure}{.24\linewidth}
		\centering
		\includegraphics[height=1.6cm]{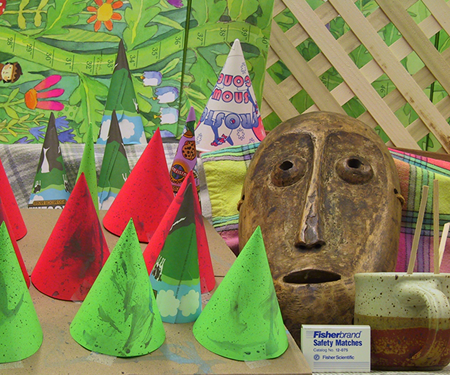}
	\end{subfigure}%
	\hspace{0.35mm}%
	\begin{subfigure}{.24\linewidth}
		\centering
		\includegraphics[height=1.6cm]{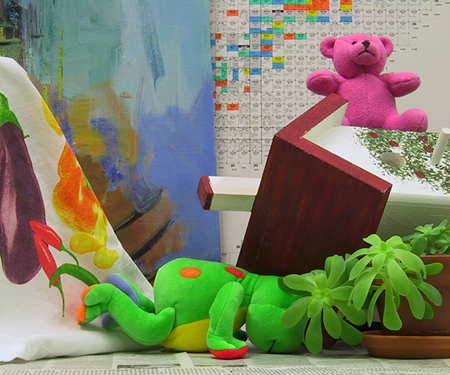}
	\end{subfigure}%
	\hspace{0.6mm}%
	\begin{subfigure}{.24\linewidth}
		\centering
		\includegraphics[height=1.6cm]{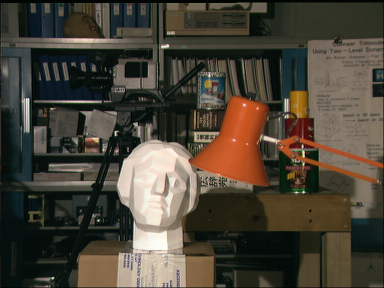}
	\end{subfigure}%
	\hspace{1.35mm}%
	\begin{subfigure}{.24\linewidth}
		\centering
		\includegraphics[height=1.6cm]{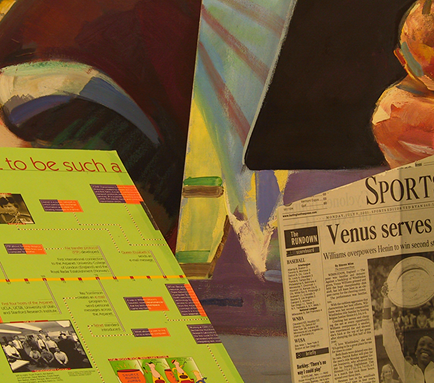}
	\end{subfigure}%
	\\[0.1cm]
	\begin{subfigure}{.24\linewidth}
		\centering
		\includegraphics[height=1.6cm]{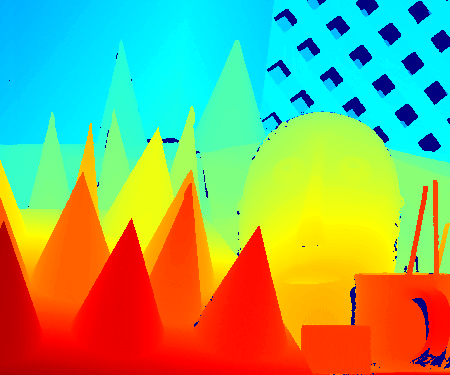}
	\end{subfigure}%
	\hspace{0.35mm}%
	\begin{subfigure}{.24\linewidth}
		\centering
		\includegraphics[height=1.6cm]{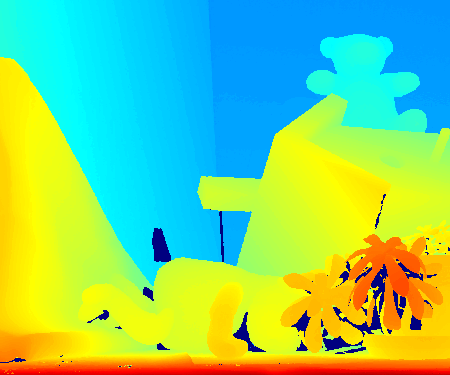}
	\end{subfigure}%
	\hspace{0.6mm}%
	\begin{subfigure}{.24\linewidth}
		\centering
		\includegraphics[height=1.6cm]{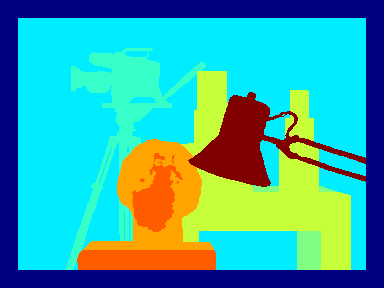}
	\end{subfigure}%
	\hspace{1.35mm}%
	\begin{subfigure}{.24\linewidth}
		\centering
		\includegraphics[height=1.6cm]{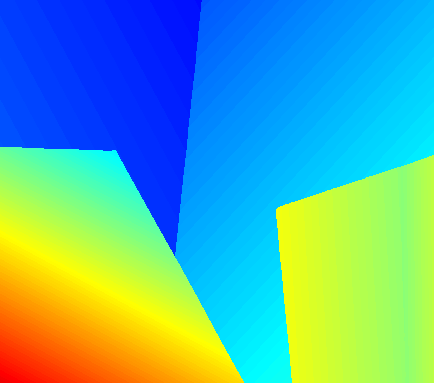}
	\end{subfigure}%
	\\[0.1cm]
	\begin{subfigure}{.24\linewidth}
		\centering
		\includegraphics[height=1.6cm]{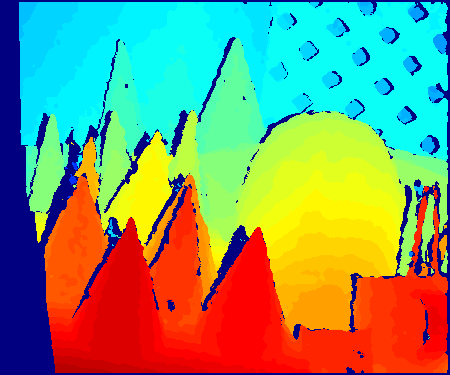}
	\end{subfigure}%
	\hspace{0.35mm}%
	\begin{subfigure}{.24\linewidth}
		\centering
		\includegraphics[height=1.6cm]{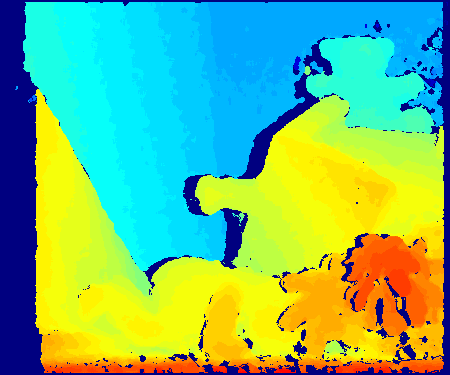}
	\end{subfigure}%
	\hspace{0.6mm}%
	\begin{subfigure}{.24\linewidth}
		\centering
		\includegraphics[height=1.6cm]{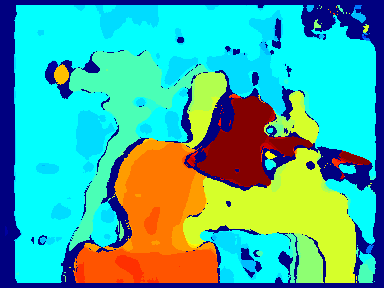} % tsukuba_our
	\end{subfigure}%
	\hspace{1.35mm}%
	\begin{subfigure}{.24\linewidth}
		\centering
		\includegraphics[height=1.6cm]{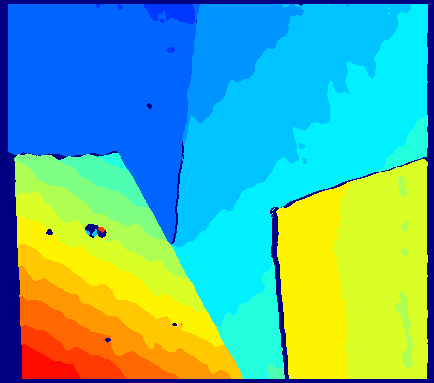}
	\end{subfigure}%
	\caption{Our results on the four most commonly used Middlebury images \cite{scharstein2002taxonomy,scharstein2003high}. Left to right: \emph{Cones}, \emph{Teddy}, \emph{Tsukuba}, \emph{Venus}. Top row: input images; middle row: ground truth disparities; bottom row: the disparities produced by our approach.}
	\label{fig:qualitativeresults-middlebury}
	\vspace{-.5\baselineskip}
\end{stusubfig}
%---

The FPGA accelerators are designed and implemented through Xilinx's High Level
Synthesis (HLS) tool, whose C++ abstraction to FPGA development allows for
faster prototyping during the design process, greater flexibility as well as
improved reusability of the resulting hardware blocks. Through the use of
Xilinx's SDSoC tool, we deploy the accelerators on the ZC706 development board.
In \autoref{tbl:quantitativeresults-windowsize}, we highlight how the frame rate
of our system is independent of the window size of the Census Transform that we
employ. Varying these parameter has, instead, an effect on the quality of the
estimated disparities: as the window size increases, the error rate on the KITTI
dataset images \cite{Menze2015CVPR,Menze2015ISA} decreases. As in
\autoref{tbl:quantitativeresults-kitti}, we report the error rate of the pixels
surviving LR check (\ie the output of the proposed method) together with their
density, and the error rate after an interpolation step done according to the
KITTI protocol. As expected, variations in the CT window size also affect the
FPGA resource utilisation of the system, \ie the number of logic/memory units
that are required to implement the necessary hardware blocks. This resource
utilisation, in turn, impacts the overall amount of power consumed by the FPGA
chip, as estimated by the Xilinx Vivado tool. This is also shown, in the last
row of \autoref{tbl:quantitativeresults-windowsize}. A justification for the
frame-rates we achieve is that the embedded system manages to output one output
disparity (for both left and right images) per three clock cycles (FPGA clocked
at 100MHz). Although the system is fully pipelined, a strict dependency is
incurred due to the use of the energy term of the previous pixel's disparity
result for the computation of the current pixel's energy values, as denoted in
Equation~\eqref{eq:ours-optimised}. More specifically, the system's bottleneck
and upper limit on frame-rate is due to the propagation delay required to
compute the minimum of any given pixel's cost vector.

\section{Conclusion}

\noindent In this paper, we have presented R$^3$SGM, a variant of the well-known
Semi-Global Matching (SGM) method \cite{hirschmuller2008stereo} for stereo
disparity estimation that is better-suited to raster processing on an FPGA. We
draw inspiration from the recent More Global Matching work of Facciolo \etal
\cite{facciolo2015bmvc}, which mitigated the streaking artifacts that afflict
SGM by incorporating information from two directions into the costs associated
with each scan line. Due to the memory access pattern involved in some of its
directional minimisations, however, MGM proves difficult to efficiently
accelerate. Instead, we propose a method that uses only a single,
raster-friendly minimisation, but one that incorporates information from four
directions at once.

%---
\begin{table}[!t]
	\centering
	\scriptsize
	\begin{tabular}{ccccccc}
		\toprule
		\textbf{Window Width} & \textbf{3} & \textbf{5} & \textbf{7} & \textbf{9} & \textbf{11} & \textbf{13} \\
		\midrule
		\textbf{Frame Rate}          & 72   & 72   & 72   & 72   & 72   & 72   \\
		\textbf{Error \%}            & 9.3  & 6.7  & 5.8  & 5.4  & 5.0  & 4.8  \\
		\textbf{Density \%}          & 73   & 81   & 83   & 84   & 85   & 85   \\
		\textbf{Interp.\ Error \%}   & 19.4 & 13.6 & 12.0 & 11.1 & 10.5 & 9.9  \\
		\textbf{LUT Utilisation \%}  & 33.4 & 37.6 & 49.9 & 58.3 & 67.3 & 75.7 \\
		\textbf{FF Utilisation \%}   & 10.8 & 14.2 & 18.1 & 24.7 & 32.2 & 40.5 \\
		\textbf{BRAM Utilisation \%} & 28.9 & 29.3 & 29.6 & 30.0 & 30.4 & 30.7 \\
		\textbf{Total FPGA Power (W)}     & 1.68 & 1.85 & 2.02 & 2.36 & 3.52 & 3.94 \\
		\bottomrule
	\end{tabular}
	\caption{
    The impact of varying the Census Transform window size on the frame rate, error and FPGA resource utilisation.
    As the window size increases, the frame rate remains constant, the error on images from the KITTI dataset \cite{Menze2015CVPR,Menze2015ISA} decreases, and the FPGA resource utilisation increases.
    Power consumption estimates were obtained from Xilinx Vivado.
    }
	\label{tbl:quantitativeresults-windowsize}
	\vspace{-\baselineskip}
\end{table}
%---

Our approach compares favourably with the two state-of-the-art GPU-based methods
\cite{kuzmin2017,hernandez2016embedded} that can process the KITTI dataset in
real time, achieving similar levels of accuracy whilst reducing the power
consumption by two orders of magnitude. Moreover, in comparison to other
FPGA-based methods on the Middlebury dataset, we achieve comparable accuracy
either at a much higher frame-rate (c.f.\ \cite{gehrig2009real,
ttofis2012towards}), using simpler, cheaper hardware (c.f.\
\cite{shan2014hardware}) or handling greater disparity ranges (c.f.\
\cite{Perri2018a, Zhang2011}). Our approach achieves a state-of-the-art balance
between accuracy, power efficiency and speed, making it particularly well suited
to real-time applications that require low power consumption, such as prosthetic
glasses and micro-UAVs.

\section*{Acknowledgements}
\noindent 
This work was supported by Innovate UK/CCAV project 103700 (StreetWise), the EPSRC grant Seebibyte EP/M013774/1 and EPSRC/MURI grant EP/N019474/1. We would also like to acknowledge the Royal Academy of Engineering and FiveAI.

{\small
	\bibliographystyle{ieee}
	\bibliography{fpt_review}
}
	
\end{document}